\documentclass[journal]{IEEEtran}

\ifCLASSINFOpdf
\else
\fi

\usepackage[cmex10]{amsmath}
\usepackage{amssymb}
\usepackage{graphicx}
\usepackage{epstopdf}
\usepackage{graphics}
\usepackage{threeparttable}
\usepackage{epsfig,color,algorithm}
\usepackage{inputenc}

\makeatletter
\def\endthebibliography{%
	\def\@noitemerr{\@latex@warning{Empty `thebibliography' environment}}%
	\endlist
}
\makeatother

\usepackage{threeparttable}
\usepackage{epsfig,color,algorithm}
\usepackage{color, colortbl}
\definecolor{LightCyan}{rgb}{0.7, 0.75, 0.71}
\definecolor{LightCyan1}{rgb}{0.53, 0.66, 0.42}
\definecolor{LightCyan2}{rgb}{0.5, 1.0, 0.83}
\definecolor{LightCyan3}{rgb}{0.98, 0.94, 0.75}
\definecolor{LightCyan4}{rgb}{0.67, 0.88, 0.69}
\newcolumntype{g}{>{\columncolor{Gray}}c}

\begin{document}

\title{Physical Layer Security for Downlink NOMA: Requirements, Merits, Challenges, and Recommendations}
\author{\IEEEauthorblockN{Haji M. Furqan\IEEEauthorrefmark{1}, Jehad M. Hamamreh\IEEEauthorrefmark{2}, and Huseyin Arslan\IEEEauthorrefmark{1} \IEEEauthorrefmark{4}}

\IEEEauthorblockA{School of Engineering and Natural Sciences, Istanbul Medipol}
\IEEEauthorblockA{\IEEEauthorrefmark{1}School of Engineering and Natural Sciences, Istanbul Medipol University, Istanbul, Turkey\\\IEEEauthorrefmark{2}Department of Electrical and Electronics Engineering, Antalya Bilim University, Antalya, Turkey\\\IEEEauthorrefmark{4}Department of Electrical Engineering, University of South Florida, Tampa, USA\\
}
}

\maketitle


\begin{abstract}
Non-orthogonal multiple access (NOMA) has been
recognized as one of the most significant enabling technologies
for future wireless systems due to its eminent spectral efficiency,
its ability to provide an additional degree of freedom for ultra
reliable low latency communications (URLLC), and grant free
random access. Meanwhile, physical layer security (PLS) has got
much attention for future wireless communication systems due to
its capability to efficiently complement the cryptography-based
algorithms for enhancing overall security of the communication
system. In this article, security design requirements for downlink
power domain NOMA and solutions provided by PLS to fulfil
these requirements are discussed. The merits and challenges
which were encountered while employing PLS to NOMA are
identified. Finally, future recommendations and prospective
solutions are also presented.
\end{abstract}
\IEEEpeerreviewmaketitle

\section{INTRODUCTION}
\IEEEPARstart{N}on-orthogonal multiple access (NOMA) has received significant attention for 5G and beyond wireless systems due to its unique properties such as high spectral efficiency, low latency, improved coverage, massive connectivity, fairness and so on \cite{Nomasurve}. However, compared to orthogonal multiple access (OMA), there are some critical security risks in NOMA. More specifically, due to the broadcast of superimposed messages from multiple users at the same time over the same resources, there is a risk that an eavesdropper can overhear the information of multiple users if NOMA transmission is successfully intercepted. Moreover, in NOMA, there is a need of securing confidential messages from each other in case of untrusted users \cite{oursur1}.        

To cope up with these security risks, physical layer security (PLS) techniques have emerged as a promising solution that can complement and (in some cases) may even replace the cryptography-based approaches \cite{oursur1} \cite{7483858}. PLS exploits the dynamic features of wireless communications, for example, random channel, fading, interference, and noise, etc., to prevent the eavesdropper from decoding data while ensuring that the legitimate user can decode it successfully. PLS approaches can be exploited to extract keys from the channel, thus avoiding key management issues. Furthermore, in PLS, channel-dependent resource allocation and link adaptation can be designed to provide flexible and scenario-specific security for 5G and beyond \cite{oursur1}. 

Based on the potential of PLS for future networks and security concerns in NOMA, designing PLS techniques for NOMA is a promising area of research. However, there is still a paucity of research works in this direction \cite{8335290}\cite{7812773}. In this article, we first provide a quick overview of NOMA flavors and basic principles to explain security concerns more clearly. This is followed by security design objectives and solutions provided by PLS. Then, we present the merits of PLS in NOMA as compared to OMA. Challenges of PLS in NOMA, possible solutions, and future directions are addressed in the following section. The final section concludes the article.
\section{DOMINANT FLAVORS AND SYSTEM MODEL FOR NOMA}
In this section, different types of NOMA, basic system model, and NOMA principles are presented to explain the security designs more clearly.

\begin{figure*}[h]
        \centering
        \includegraphics[width=7in,,height=2.4in]{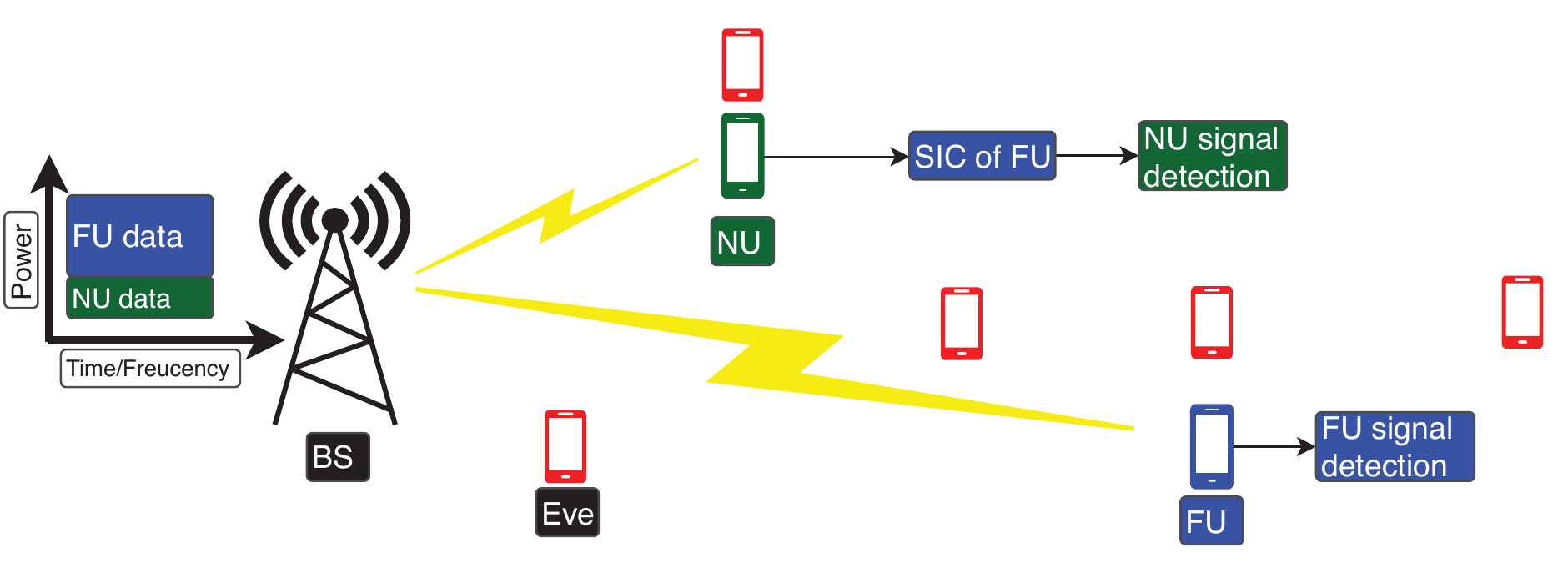}
        \caption{Downlink NOMA detailed model which consists of a single Base Station (BS) with one Near User (NU) and one Far User (FU) in the presence of an external eavesdropper (cloned at different possible positions).}
        \label{sysm}
    \end{figure*} 
\subsection{NOMA Dominant Flavors}
NOMA supports massive connectivity and enhanced spectral efficiency by allowing resource allocation in a non-orthogonal manner. There are two basic types of NOMA schemes: Power-domain (PD) NOMA  and code-domain (CD) NOMA \cite{Nomasurve}. In PD-NOMA, different users' signals are directly superimposed by assigning channel quality-based power allocation to them, while sharing the same frequency-time resources. CD-NOMA, on the other hand, is like Code Division Multiple Access (CDMA), where different users are allowed to share the same frequency-time resources by using unique orthogonal code. However, CD-NOMA uses non-orthogonal codes with lower cross-correlation or sparse sequences. While uplink NOMA \cite{8681607} has also been studied, more focus is given to downlink NOMA,
especially by the standardization bodies,  such as the third generation partnership project (3GPP) and IEEE. For example, a downlink version of PD-NOMA has been proposed for 3GPP-LTE-Advanced \cite{3GPP_38_824}. Hence, this paper will mainly focus on PLS techniques applied to downlink PD-NOMA to elaborate on the novel challenges and future recommendations for it.\footnote{Note that we will use the term “NOMA” in the remaining part of the paper to represent “PD-NOMA” \cite{Nomasurve}.}
\subsection {System Model and Principles of NOMA}
Consider a simple two-user downlink NOMA scenario that consists of a single base station (BS) with one near user (NU) and one far user (FU) in the presence of an external eavesdropper (cloned at different possible positions) as shown in Fig. \ref{sysm}. The BS first superimposes the users' signals by allocating them different power levels and broadcasts the mixture to all users using the same time-frequency resources. The power allocation in NOMA is done in such a way that the FU (user with lower channel gain) is allocated more power and NU (user with higher channel gain) is given low power. The receivers of NOMA employ different strategies for different users in accordance with their channel characteristics. More specifically, the NU has to decode the signal intended for FU first, and afterward, it subtracts the detected signal from the received signal and then decodes its intended data. This process is known as successive interference cancellation (SIC). On the other hand, the FU directly decodes its information while considering the information of its partner as noise. It should be noted that for the sake of explanation two users case is considered here; however, the discussion is also applicable to multiple (more than two) users case.
The above-mentioned case is for single-input-single-output (SISO)-NOMA, where channels are represented by scalars. However, matrices are used to represent the channels of multi-input-multi-output (MIMO)-NOMA. In the case of matrices, ordering of users based on power is quite challenging \cite{Nomasurve}. In the literature, two main designs are proposed for MIMO-NOMA case: 1) \textit{Beamformer based MIMO-NOMA}, where different beams are allocated to different users and SIC is employed at users sharing the same resource block \cite{Nomasurve}, 2) \textit{Cluster based MIMO-NOMA}, where users are divided into clusters and a single beam can serve all the users in the cluster. In this approach, SIC is adopted among users sharing the same cluster \cite{Nomasurve}. 

\begin{figure*}[t]
        \centering
        \includegraphics[width=6in,,height=2.4in]{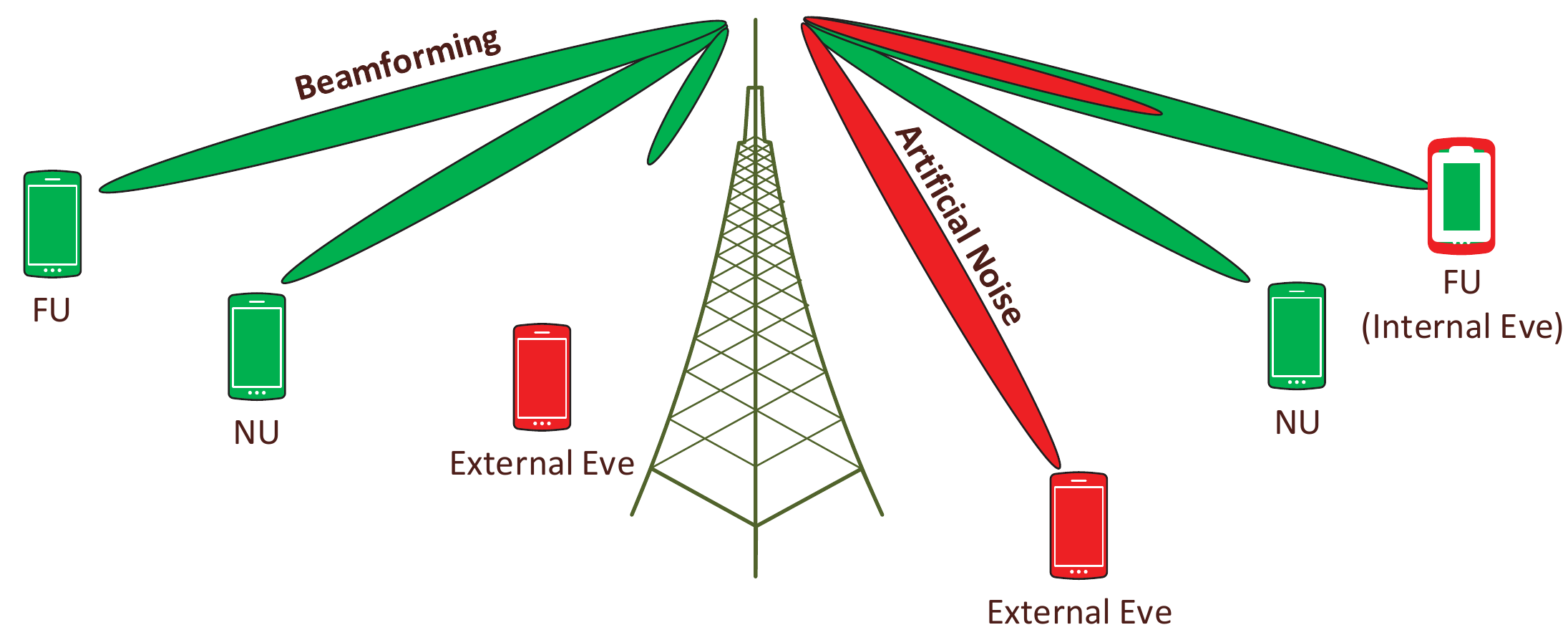}
        \caption{Security based approaches for internal and external eavesdropper based on beamforming and AN.}
        \label{beam}
    \end{figure*} 
\section{SECURITY DESIGNS OBJECTIVES}
In this section, different security design objectives for NOMA are presented and explained. To evaluate the secrecy performance of any security algorithm, the secrecy rate is one of the popular metrics in PLS. It is defined as the difference between the capacity of the legitimate user channel (main channel) and the capacity of the eavesdropper channel (wiretap channel). 

In general, different users in NOMA can have different requirements in terms of reliability, throughput, and security, etc. which implies that the design of PLS techniques should consider these requirements. Moreover, there are two types of eavesdroppers: 1) {External}, and 2) {Internal}. An internal eavesdropper is from the set of legitimate users of the network, while the external one is not from that set \cite{7483858}. The eavesdropper can be considered 1) active, or 2) passive. The active eavesdropper can interrupt wireless communication by launching jamming or channel estimation attacks while passive eavesdropper just spies on the communication without interfering with the ongoing communication.

This work is focused on both internal eavesdropping as well as passive external eavesdropping. The objectives of security design for NOMA can be divided into three major categories based on its requirements as follows:
\begin{itemize}
\item
Security designs against external eavesdroppers.
\item
Security designs against internal eavesdroppers.
\item
Security designs against both internal and external eavesdroppers.
\end{itemize}
The details of security designs and solutions provided by PLS are presented in the subsequent part.
\subsection{Security Designs against External Eavesdroppers}
In this scenario, NU and FU are trusted. So, the design goal here is to secure the messages of NU and FU from an external eavesdropper. Based on the basic model presented in Fig.\ref{sysm}, there are different possibilities for Eve’s location. In some cases, Eve is closer to BS compared to users, and her channel is better than far legitimate users. Hence, the location of Eve can affect the security performance of NOMA system and should be considered while designing security algorithms. Moreover, different users are allocated different power levels, due to which they are protected in an unequal manner with respect to Eve's location. More specifically, Eve can eavesdrop their signals to different extents.

The necessary conditions that need to be taken into consideration while designing algorithms for this scenario are as follows: \textbf{Firstly}, the basic SIC should be normally operated with the security algorithm, which means that the proposed algorithm should not affect the basic SIC process and the performance of normal NOMA. \textbf{Secondly}, the algorithm is also expected to work even in the case of having strong spatial similarity between channels of legitimate and illegitimate parties.

The popular PLS techniques for external eavesdropping in the literature include channel-based optimization of the power allocation for each user, subcarrier assignment to users, channel ordering of NOMA users along with the decoding order, optimization of beamforming policies, adding interfere signal, key generation, phase manipulation, transmit antenna selection (TAS) approaches and inter-user interference exploitation, etc. \cite{oursur1}\cite{8335290}. The brief details of popular approaches are as follows:
\subsubsection{Beamforming}
The basic idea of the beamforming-based security approach in OMA is to enhance the power of the signal at the legitimate users while suppressing it in other directions \cite{oursur1} as presented in Fig. \ref{beam} (left). However, this approach may not be able to fulfill the above-mentioned design requirements for secure NOMA. For example, the beamforming design matrices based on maximum ratio transmission for near and far users increase the strength of both users' signals which may not guarantee perfect SIC processing at near user \cite{feng2019beamforming}. Hence, these techniques need to be intelligently modified.
\subsubsection{Artificial noise (AN) with beamforming}
AN based techniques with beamforming are very effective against external eavesdropping in NOMA, especially when Eve is closer to BS compared to the legitimate user. The basic idea is to transmit intentional interference simultaneously with the desired signal by using the beamforming approach to degrade the performance of eavesdropper while fulfilling the above mentioned basic security design requirements for NOMA as presented in Fig. \ref{beam} (right bottom). The performance of such types of techniques is highly dependent on the availability of channel state information (CSI) of the eavesdropper. In the case of full CSI availability at the BS, optimal and efficient beamformers can be designed to enhance the security \cite{8335290}. However, when CSI is not available, the beamformer should be designed to send AN in all directions except in the direction of the desired user while sending the intended signal in the direction of the desired user \cite{feng2019beamforming}. The major challenge here is to ensure secure communication while fulfilling the above-mentioned conditions.
\subsubsection{Power allocation}
Power allocation approaches based on channel conditions of legitimate users can make the interception of users’ signals difficult for eavesdropper under certain settings in NOMA \cite{7426798}. In the case of full CSI availability, the power allocation can be optimized to maximize the secrecy rate (security) of the legitimate users \cite{7426798}. However, in the case of imperfect CSI, optimal power allocation for maximizing secrecy rate (security) is not possible \cite{8335290}. So, in such cases, the goal is to maximize the difference in data rate between Eve and users as much as possible.
\subsubsection{Cooperative beamforming and jamming}
Cooperative communication can enhance the reliability of NOMA systems by cooperative diversity. Moreover, it can also enhance the security of the NOMA system by distributed beamforming with and without cooperative jamming. In the case of distributed beamforming, the signal is directed towards the desired direction by collaborative action of relays \cite{arafa2019secure}. On the other hand, in case of distributed beamforming with cooperative jamming, a group of relays is selected to focus the desired signal in the intended direction while the remaining relays are used to degrade the performance Eve by sending AN \cite{arafa2019secure}. 
\subsection {Security Designs against Internal Eavesdroppers}
In this scenario, no external eavesdropper is assumed; however, the users are untrusted. The design goal here is to secure information of users from each other, while making sure that the SIC operation works normally. Moreover, in this case, the channel is known at the BS, which makes the design process different than the previous case. Internal eavesdropping can be divided into two types:
\begin{itemize}
\item
Eavesdropping of FU by NU
\item
Eavesdropping of NU by FU
\end{itemize} 
\subsubsection {Eavesdropping of FU by NU}
In the basic NOMA principle, the main security risk for FU is that the NU has to decode (or demodulate) the signal of FU in order to apply SIC. Another important thing is that the FU's signal is allocated more power, which makes its detection easier for the NU. The design goal here is to avoid leakage of information of FU to NU, while making sure that SIC works normally. To further elaborate on this issue, it should be pointed out that there are two types of SIC receiver: The first one is \textbf{symbol-level SIC receiver}, in which FU’s signal is demodulated but not decoded in order to apply SIC, while the other one is \textbf{codeword-level SIC receiver}, where FU's signal is demodulated and decoded in order to apply SIC. In the codeword-level-SIC case, the data can only be secured by cryptography-based techniques. However, for the case of symbol-level-SIC, PLS techniques can be applied. In symbol-level based SIC, security can be provided to FU’s data by transforming its data into another domain by using a special sequence such that NU can apply SIC normally, but cannot decode the information of FU \cite{7997115}. Moreover, this transformation can also be done by using channel-dependent features. Note that there are not too many contributions to the literature in this direction.
\subsubsection {Eavesdropping of NU by FU}
In the basic NOMA principle, the FU can decode its signal directly, considering the information of near as noise. However, after obtaining its own signal, it may detect the signal of NU. The design goal here is to secure the data of NU from FU while making sure that SIC works normally. In this case, designing security methods is easier as compared to the security problem of FU’s data. The BS can employ PLS techniques based on power allocation, beamforming, or any other adaptation-based algorithm to satisfy the security requirement of NU while making sure that the basic data rate requirement of FU is fulfilled. For example, in the case of beamforming, the design should consider the balance between ensuring security at the near user while reliability at the far user, as presented Fig. \ref{beam} (right top). 
\subsection {Security Designs against both Internal and External Eavesdroppers}
In this scenario, there is an external eavesdropper as well as an internal eavesdropper where the users in the network are not trustable. The design goal here includes the security of signals intended for NU and FU from external eavesdropper as well from each other. This case is the most challenging one with respect to security design. The design algorithms should make sure that SIC will work normally while fulfilling the above goals. One possible way to provide security, in this case, can be by the transformation of the signal of near and far users into another domain by using some randomization sequences \cite{7997115}. However, this is still an open research area, and a lot of research efforts are needed in this direction.
A summary of the objectives of security designs, complexity, and popular solutions for NOMA are presented in Table. I.

 \begin{table*}[t]
  	\begin{center}
  		\caption{Summary of the objectives of security designs for different scenarios in NOMA focusing on passive External Eavesdropper and Active internal eavesdropper.}
  		\label{T13}
  		{\small
  			
  		\begin{tabular}	
   {|p{.17\textwidth}|p{.25\textwidth}|p{.15\textwidth}|p{.24\textwidth}|}
  						\hline  
    \rowcolor{LightCyan}
  	\textbf{Scenarios for security} & \textbf{Design objectives}& \textbf{Design Complexity}&\textbf{Candidate solutions}\\  \hline
  		 \rowcolor{LightCyan4}
  		{External Eavesdropper} & {Securing NU and FU data against external Eavesdropper while keeping normal SIC
 }& {Normal}&{Beamforming, Power allocation based,
 interference exploitation based, TAS, relay selection, etc.
 }\\  \hline
  	\rowcolor{LightCyan2}
 		{Internal Eavesdropper} & {Securing users' information from each other while ensuring normal SIC} & {\textbf{Against NU}: High \newline \textbf{Against FU}: Normal}
 &{\textbf{Against NU}:
 Transformation of FU to other domain \newline
 \textbf{Against FU}:
 Beamforming
 Power allocation, TAS etc.}
 \\  \hline
  	\rowcolor{LightCyan3}
  \rowcolor{LightCyan3}
  	{External and \newline Internal Eavesdropper
 } & {Securing users' information from each other as well as from external Eve while having normal SIC
 }& {Highest
 }&{Transformation of users' signal into another domain, channel based phase rotation, interference assisted, etc.
 }\\  \hline
  			\end{tabular}
  		}
  	\end{center}
  \end{table*}
 
\section{MERITS OF PLS IN NOMA}
In this section, we present some of the merits of NOMA over OMA with respect to PLS under certain conditions. 
\subsection{Higher Sum-Secrecy Rate}
\begin{figure}[h]
        \centering
        \includegraphics[width=3.5in,,height=2.5in]{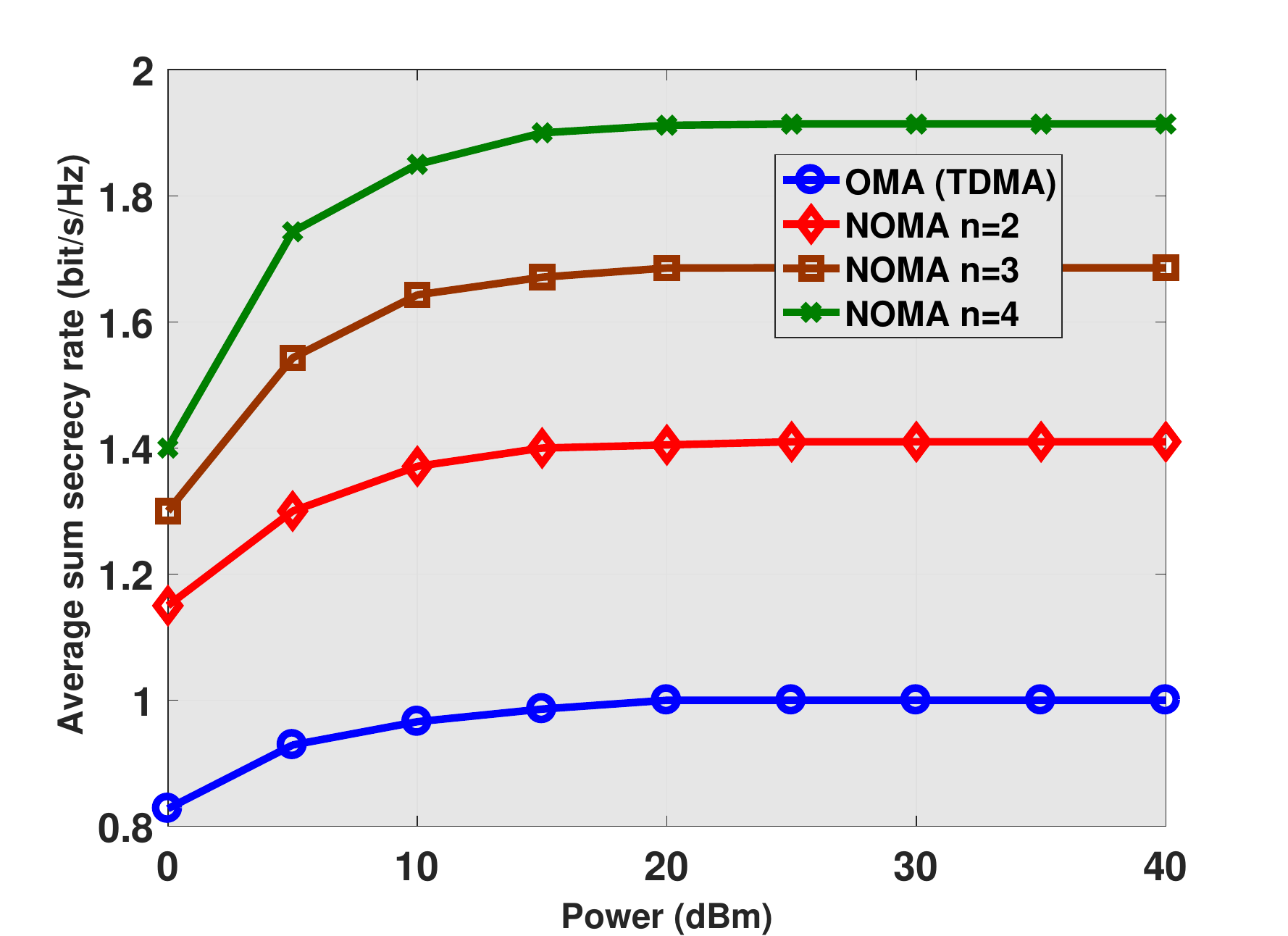}
        \caption{Average Sum Secrecy Rate (ASSR) versus the transmit power for different number of users (\textit{n}=2, \textit{n}=3, \textit{n}=4).}
        \label{SSR}
    \end{figure}

In NOMA, the signals are not sent separately like OMA. Hence, multi-user interference and PLS can be processed collaboratively. Moreover, user selection, number of clusters, intra-cluster and inter-cluster power allocation can be designed based on the quality of service requirements of legitimate users such as data rate, reliability, etc. to enhance the secrecy rate of the system. For example, power allocation based on channel conditions of legitimate users can enhance the security of the system under certain settings as presented in Fig. \ref{SSR} of \cite{7426798} . It should be noted from the figure that the average sum secrecy rate (ASSR) of the NOMA system improves with the increase in the number of users as compared to OMA under specific settings. The reason for the improvement in ASSR is due to the dominance of legitimate users' high spectral efficiency \cite{7426798}.
\subsection{Inter-User Interference Exploitation for Securing Massive MIMO System}
\begin{figure}[h]
        \centering
        \includegraphics[width=3.21in,,height=2.5in]{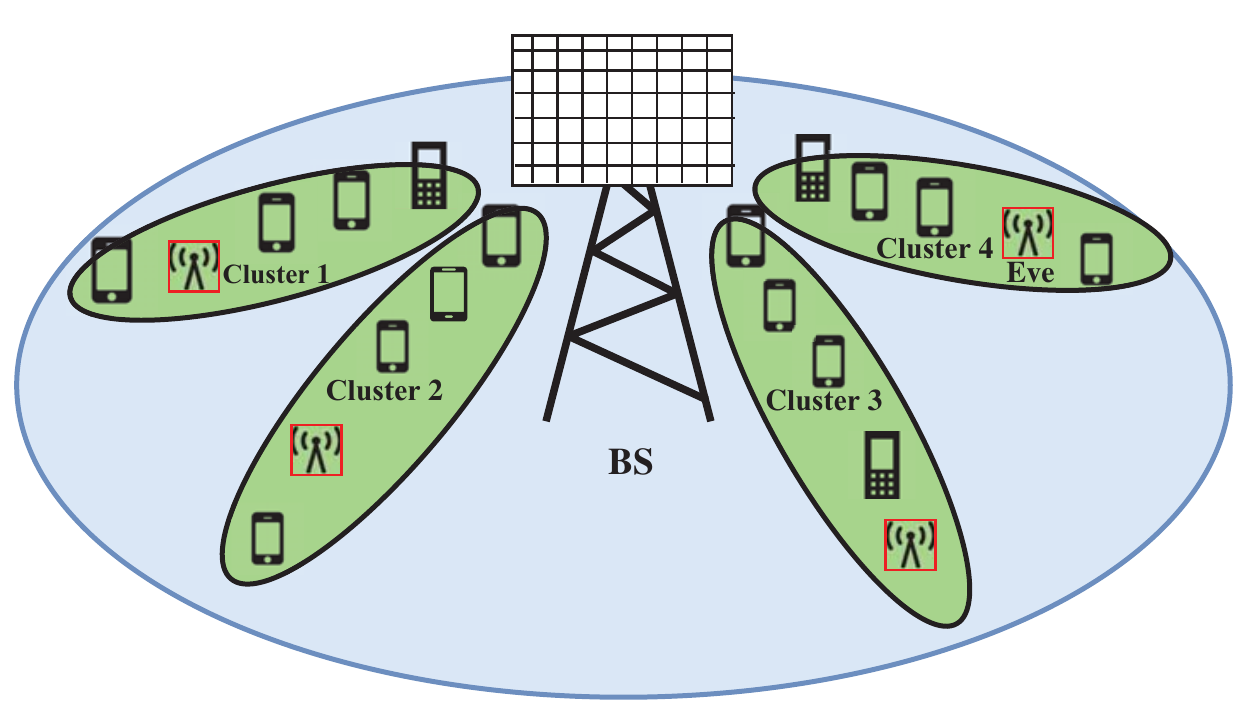}
        \caption{Secure massive MIMO with NOMA by using inter-user interference, where users are divided into four clusters \cite{8334223}.}
        \label{interuser}
    \end{figure}

In the case of a massive MIMO system, AN-based security techniques face complexity issues. In such cases, NOMA can help us to provide secure communication without using AN \cite{8334223}. For example, consider a clustering-based Massive MIMO NOMA system employing non-orthogonal channel estimation in the presence of multiple active eavesdroppers \cite{8334223} as presented in Fig. \ref{interuser}. The nodes in this system suffer from intra-cluster and inter-cluster interference; however, this inter-user interference can be exploited intelligently in NOMA to provide secure communication \cite{8334223}. More specifically, power allocation coefficients during channel estimation and multiple access stages can be designed in such a way that it will enhance the performance of legitimate users and degrade the performance of active eavesdroppers \cite{8334223}. Moreover, this approach can also be extended to full-duplex NOMA to provide secure communication \cite{8335290}.

\subsection{Securing Uni-Cast Message from Multi-Cast Receivers}
\begin{figure}[h]
        \centering
        \includegraphics[width=3in,,height=2.5in]{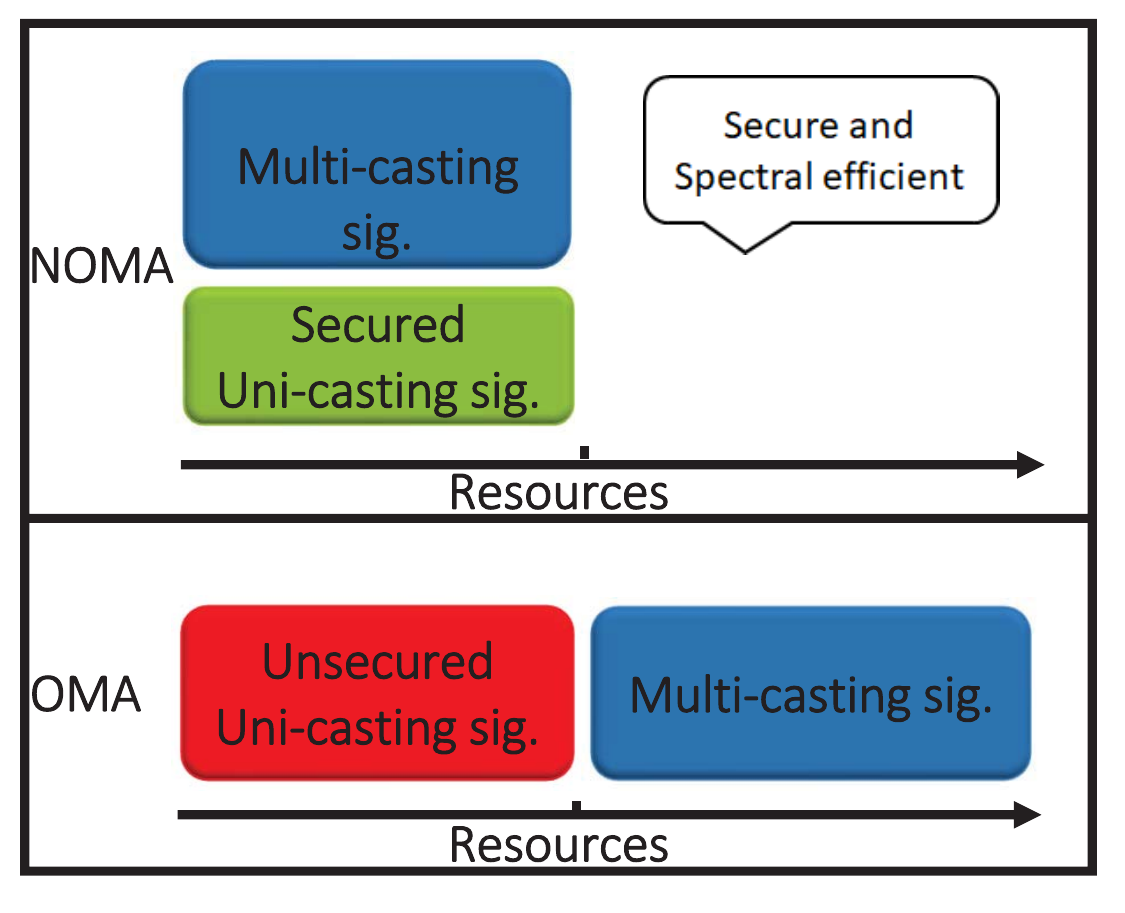}
        \caption{Multi-casting and Uni-casting in NOMA and OMA \cite{7906532}.}
        \label{securecast}
    \end{figure}

An interesting advantage of NOMA is to secure a uni-cast message from interception by the untrusted multi-cast receivers while improving spectral efficiency \cite{7906532} as presented in Fig. \ref{securecast}, where uni-cast message is for a specific receiver while the multi-cast message is for all the receivers in the set of specific receivers. In OMA, uni-casting and multi-casting are transmitted separately and can be intercepted easily by multi-casting receivers as presented in Fig. \ref{securecast}. However, the NOMA principle can be used to degrade the intercepting capabilities of the multi-casting receivers similar to the case of securing NU message from untrusted FU receiver \cite{7906532}. More specifically, joint power allocation and beamforming strategies can be used to enhance the secrecy of uni-cast message while preserving the reliability of multi-cast message \cite{7906532}. Moreover, in OMA, two slots are required to send uni-casting and multi-casting information while in NOMA both information types can be transmitted simultaneously by using a single slot \cite{7906532} as presented in Fig. \ref{securecast}. 
\subsection{Channel Correlation and Security}
Most, if not all, PLS techniques (based on small scale fading) assume that the received signals at Eve and Bob will experience independent fading if they are roughly half a wavelength apart. This assumption is valid only in a sufficiently rich scattering environment. In the case of a poor scattering environment, these algorithms will not ensure secure communication. However, NOMA with large scale fading based security algorithms can provide secure communication under certain circumstances even in a poor scattering environment \cite{7812773}\cite{8334223}. Moreover, in the case of a rich scattering environment, both the small and large scale fading based security algorithms can be applied in NOMA.
\section{CHALLENGES AND FUTURE RESEARCH DIRECTIONS}
This section presents the challenges in securing NOMA using PLS alongside some of the proposed solutions and research directions.
\subsection{Challenges for Security against FU and External Eavesdroppers}
There are considerable contributions in the literature regarding the provision of secure communication schemes against untrusted FU and external eavesdroppers (in case of trusted internal users), such as channel-dependent power allocation, beamforming, cooperative communication, TAS and inter-user interference, etc.. However, the majority of the research works assumes the availability of full, partial or statistical information about the CSI of Eve, which is difficult to achieve in case of passive eavesdropping. Moreover, some techniques provide security at the cost of performance degradation. Hence, conventional techniques should be intelligently modified, and novel techniques should be proposed to provide security for NOMA. Some of the potentially interesting techniques like multi-dimensional directional modulation scheme, cyclic feature suppression-based techniques, and channel-based interleaving, etc., have not been explored for such cases, whereas these techniques have the potential to be used in such situations \cite{8335290}. 

\subsection{Security Challenges against Untrusted NU and both External and Internal Eavesdroppers}
The design of security algorithms against untrusted NU and both internal and external eavesdropper is extremely challenging. The only solutions available in the literature so far are based on the transformation of signals into another domain. 
This transformation is done by using a transformation sequence that needs to be shared between the legitimate parties \cite{7997115}. The sequence can be shared by PLS approaches, such as full-duplex jamming based techniques for sequence sharing \cite{7997115} which requires complex hardware. In this direction, the channel-based phase manipulation of symbols, directional modulation and cross-layer security techniques can also be effective. For example, automatic repeat request (ARQ) with AN can be jointly designed to provide security against internal and external eavesdropping in NOMA similar to the work presented in \cite{8415757}. Moreover, joint composite constellation design and ARQ with adaptive modulation can also be used to provide security against untrusted NU. Furthermore, in the case of a rich scattering environment, channel-based manipulation security techniques can also be employed in such scenarios. This is still an open area and a lot of research efforts are needed to provide security for such cases while making sure the SIC operation works normally.
\subsection{Passivity and Limited Observations}
A lot of techniques in the literature of secure NOMA consider that the illegitimate user is just spying the information. However, in future networks, there may exist illegitimate nodes that can interfere with the normal operation of the NOMA system by active attacks, such as pilot spoofing attacks, etc. These attacks are more critical in NOMA because of the broadcast of superimposed messages of multiple users at the same time.
Quite a few PLS techniques in the NOMA literature are robust to active eavesdroppers’ case \cite{8334223}. Hence, there is a need of designing PLS techniques that are robust to active attacks from eavesdroppers. Moreover, collaborative-eavesdroppers with multiple observations may lead to zero secrecy rate \cite{7483858}. Hence, there is a need for understanding the implications of collaborative-eavesdroppers and multi-observation cases while developing security techniques for NOMA.

\subsection{SIC and Eve Capability}
In the literature, it is assumed that eavesdroppers use the same SIC procedure as legitimate users. However, an eavesdropper can apply alternative strategies for eavesdropping, for example, it may decode a signal in the first step that is decoded in the last stage of SIC at legitimate users, which can affect the overall security performance of the system. Moreover, a powerful eavesdropper can apply parallel interference cancellation to simultaneously decode the users' signal \cite{Nomasurve}. Possible alternative approaches by eavesdropper should also be considered while designing security algorithms. 
\subsection{SIC Error Propagation and Secrecy}
The security algorithms in NOMA mainly rely on the assumption that perfect channel estimate is available, and the signals are perfectly separated at the receiver side (perfect SIC). However, if there is an error in any of these signals during SIC, then the remaining signals may also be detected erroneously \cite{Ding2018}. Hence, the effect of imperfect SIC and imperfect channel estimation should be considered while designing security algorithms for NOMA, so that these drawbacks can be avoided. Therefore, it is also recommended to use an efficient non-linear detection algorithm at each state of SIC to alleviate the effect of imperfect SIC and practical channel calibration solutions for imperfect channel estimation case. Moreover, new interference cancellation schemes and improvement in signal processing chip technology that can benefit the legitimate receivers are also of special interest \cite{Nomasurve}.

\subsection{AN based Security Schemes}
AN-based techniques are one of the popular techniques in the literature. In these techniques, an artificial interference signal is added in the null-space of the legitimate user channel to degrade the performance of Eve. However, in NOMA, when AN is added based on the individual user, it also causes AN leakage in the range space of other NOMA users which degrades their performance. Moreover, AN may increase peak to average power ratio (PAPR), sacrifices some power, and is also sensitive to imperfect channel estimation. Thus, it is recommended to design AN, not only to provide security but also to reduce the amount of out-of-band emission (OOBE), adjacent channel interference and average PAPR, etc. \cite{8415757}.

\subsection{Multi-Cell Case and Other Technologies}
In the case of multi-cell NOMA, there are a lot of challenges to provide secure and reliable communication due to inter-channel interference. However, there is not much work in this area. Algorithms for joint processing, coordinated beamforming, and coordinated scheduling need to be proposed to ensure reliable and secure multi-cell NOMA.
Moreover, there is also the paucity of PLS research works for NOMA integrated with other technologies such as millimeter-wave, full-duplex, visible light communication, cognitive radio, heterogeneous networks, and coordinated multi-point, etc.

\subsection{Cross-layer, Context-Aware and Hybrid Security Techniques for NOMA}
In the literature of PLS techniques in NOMA, transmission parameters of the physical layer are optimized according to legitimate users' channel characteristics to provide secure communication without considering upper layer parameters. However, to meet the diverse requirements of NOMA users and for joint design of throughput, secrecy, delay, reliability, and respective trade-off among them, the concept of cross-layer security design from the perspective of physical layer should also be considered such as: 1) Cross MAC-PHY layer: In this approach, MAC layer features (for example, channel accessing, multiplexing, ARQ and control of resource allocation, etc.) can be optimized jointly with physical layer parameters to provide efficient QoS based security solution \cite{8415757}, 2) Cross NET-PHY layer security: In this approach, the network layer features such as relaying, routing and path determination, etc. can be optimized jointly with physical layer parameters for enhancing security of the system \cite{7483858}, 3) Cross APP-PHY layer: In this approach, physical layer parameters of transmission are jointly optimized based on channel characteristics as well as on the basis of applications, services and features of data to provide efficient security solution based on the requirements of users.
Finally, designing \textbf{hybrid} techniques by combining signal’s security approaches (PLS) with data security approaches (cryptography) can further enhance the security of the NOMA-based systems.
\subsection{IRS assisted PLS for NOMA}
Recently, reconfigurable intelligent surfaces (RIS)-assisted networks have been proposed as a promising power-efficient solution to enable a smart and controllable wireless propagation environment. Basically, the RIS is a large array of passive reflecting elements that intelligently reflect the impinging signals in order to add different signals constructively or destructively at receivers \cite{8910627}. This feature can be exploited to enhance PLS against external and internal eavesdropper in NOMA \cite{8910627ss}.
\section{Conclusion}
NOMA promises high spectral efficiency, low latency, and massive connectivity, while PLS offers simple and effective security solutions. Together, these two technologies are capable of supporting the exceeding efficiency and security requirements of 5G and beyond networks. In this article, the key security design requirements of NOMA and the strength of PLS as a solution to fulfill these requirements are discussed. By employing PLS to NOMA, spectrally efficient, adaptive, and secure systems can be realized. However, the challenges and future recommendations explained in this work need to be investigated further to address the open issues. Practical secure NOMA systems can be developed by modification of current PLS techniques and/or proposing new novel techniques that do not require extra processing, extra signaling, or major modification in the receiver structure.

\end{document}